\documentclass[11pt]{article}

\title{\textbf{Experiential Versus Instructional Approaches for Eliciting Metacognitive Awareness in AI-Assisted Learning: A Short-Term Longitudinal Study}}

\author{
    \fontsize{11}{13}\selectfont 
    1\ts{st} Pau Benazet i Montobbio \\
    \fontsize{10}{11}\selectfont 
    Universitat Pompeu Fabra\\
    \fontsize{10}{11}\selectfont 
    Barcelona, Spain\\
    \fontsize{10}{11}\selectfont   pau.benazeti01@estudiant.upf.edu\\
    \fontsize{10}{11}\selectfont 
    ORCID: 
    \href{https://orcid.org/0009-0007-7759-6016}{0009-0007-7759-6016}
    \and
    \fontsize{11}{13}\selectfont 
    2\ts{nd} Janne Rotter \\
    \fontsize{10}{11}\selectfont 
    Universitat Pompeu Fabra\\
    \fontsize{10}{11}\selectfont 
    Barcelona, Spain\\
    \fontsize{10}{11}\selectfont   jannedavid.rotter01@estudiant.upf.edu\\
    \fontsize{10}{11}\selectfont 
    ORCID: 
    \href{https://orcid.org/0009-0001-3520-958X}{ 0009-0001-3520-958X}
    \and
    \fontsize{11}{13}\selectfont 
    3\ts{rd} Davinia Hern\'{a}ndez-Leo \\
    \fontsize{10}{11}\selectfont 
    Universitat Pompeu Fabra\\
    \fontsize{10}{11}\selectfont 
    Barcelona, Spain\\
    \fontsize{10}{11}\selectfont   davinia.hernandez-leo@upf.edu\\
    \fontsize{10}{11}\selectfont 
    ORCID: 
    \href{https://orcid.org/0000-0003-0548-7455}{0000-0003-0548-7455}
}

\date{}

\usepackage{lipsum}
\usepackage{lmodern} 
\usepackage{array} 
\usepackage{booktabs} 
\usepackage{dblfloatfix}

\usepackage[
layout=letterpaper, 
paper=letterpaper, 
portrait, 
head=0.5in,
foot=0.5in,
top=1in, 
bottom=1in, 
left=0.75in, 
right=0.75in
]{geometry}

\usepackage{tabularx}
\usepackage{booktabs} 
\usepackage{longtable} 
\usepackage{caption}
\usepackage{subcaption}

\usepackage[english]{babel}
\renewcommand{\thesection}{\arabic{section}}
\renewcommand{\thesubsection}{\thesection.\arabic{subsection}}

\usepackage{enumitem}
\setlist{itemsep=0.3em, topsep=0.3em}
\usepackage{xurl}

\usepackage[runin]{abstract}

\usepackage{hyperref}

\hypersetup{
colorlinks=true,
urlcolor=blue,
linkcolor=blue,
citecolor=blue,
pdftitle={Experiential Versus Instructional Approaches for Eliciting Metacognitive Awareness in AI-Assisted Learning: A Short-Term Longitudinal Study - Benazet i Montobbio et al.},
pdfsubject={Original Article},
pdfauthor={Benazet i Montobbio et al.},
pdfkeywords={Generative Artificial Intelligence, Metacognition, Experiential Learning, Higher Education, Human-AI Collaboration, Cognitive Offloading}
}

\usepackage[backend=biber,style=apa,sorting=nyt]{biblatex}
\usepackage{etoolbox}
\usepackage{indentfirst}

\usepackage{custom_commands} 

\usepackage{subfiles} 

\usepackage[pdftex]{graphicx} 

\begin{document}
\renewcommand{\abstractname}{}

\maketitle 


\setlength{\absleftindent}{0em}
\setlength{\absrightindent}{0em}

\begin{abstract}
    \abstractText 
\end{abstract}

\section*{Practitioners notes}
\label{chap:practioners}
\textbf{What is already known about this topic:}
\begin{itemize}
\item Experiential teaching approaches expose learners to the subject of interest and provide hands-on experimentation while instructional approaches rely on classical lecture based communication and direct instructions. There is still an active pedagogical debate about which is more beneficial for students.
\item Metacognition, which describes the ability to monitor and regulate one's own cognition, is central to learning and to the question of how to beneficially use technological tools such as AI during the learning process. 
\end{itemize}
\textbf{What this paper adds:}
\begin{itemize}
\item Compared with instruction, experiential learning produced larger immediate gains in students' knowledge of effective AI-use strategies and in their engagement with AI tools. 
\item These immediate gains were not matched by a rise in Metacognitive Regulation of Cognition, students' monitoring and adjustment of how they use AI. This raises the concern that knowing good strategies and using AI more does not by itself mean students can apply them. 
\item Over five weeks the two approaches converged on knowledge and engagement. However, the experiential group showed a delayed, continued rise in Metacognitive Regulation of Cognition, which suggests regulatory skills develop more slowly than knowledge under experiential learning. 
\end{itemize}
\textbf{Implications for practice and/or policy:}
\begin{itemize}
\item Practitioners should prioritize hands-on activities over lecture-based instruction in AI-assisted learning, as they build strategy knowledge faster and may support the slower development of regulatory behaviour. 
\item Practitioners should remain patient if experiential exercises don't produce immediate results, as these activities often need time to take effect. Instead, they should create opportunities for independent practice, allowing students to gradually strengthen their self-regulatory skills through repeated engagement and guided reflection.
\end{itemize}

\section{Introduction}
\label{chap:intro}
In modern higher education, Generative AI (GenAI) has become deeply embedded in students' daily lives and learning habits. It not only allows students easy access in natural language to the combined resources of the web but also assists in generating and refining writing and ideas \parencite{Kasneci2023}. A large-scale survey of nearly 3,800 university students spanning 16 countries revealed that by 2024, 86\% of students reported actively incorporating AI into their academic work \parencite{DigitalEducationCouncil2024}. As students are already using these tools regardless of institutional policies \parencite{Yan2024}, the focus of many scholars and practitioners has shifted from whether to allow AI toward how to effectively integrate it with students' learning. 

While AI-assisted learning offers many advantages, such as personalized support, timely feedback, and adaptive learning material \parencite{Giannakos2025, Yan2024}, a number of equally important challenges arise. These include concerns about bias in model output, accountability, privacy, and the impact on students' cognitive and personal development \parencite{Chan2023, Mittal2024, Yan2024}. Especially when used unreflectively, AI was shown to hinder learning by negatively impacting decision-making, critical thinking, and analytical reasoning \parencite{Zhai2024}, which is particularly worrisome considering the high usage patterns among university students.

\subsection*{The Metacognitive Challenge}

These findings highlight the need to carefully scaffold AI-assisted learning in higher education contexts. The challenge is fundamentally metacognitive, a concept popularly defined as “thinking about one's own thinking” \parencite{Flavell1979}. Metacognition during learning was shown to even outweigh factors like intelligence \parencite{Veenman2005}, with some researchers arguing it distinguishes good from less effective learners \parencite{Jaleel2016, Mahdavi2014}. While many different frameworks for this term exist, the concept is commonly split into two components \parencite{Schraw1994}. Metacognitive Knowledge of Cognition (MKC) refers to what learners know about their own cognition and strategies, encompassing declarative knowledge (knowing that), procedural knowledge (knowing how), and conditional knowledge (knowing when and why) (i.e. "What do I know about how I learn?"). Metacognitive Regulation of Cognition (MRC) refers to the processes through which learners control that cognition, namely planning, monitoring, and evaluating their own thinking and adjusting it accordingly (i.e. "How should I adjust my thinking?"). 

Both components are crucial when considering how students learn with GenAI. Metacognitive Knowledge of Cognition gives students an understanding of how GenAI use can support or undermine their learning, while Metacognitive Regulation of Cognition is what allows them to monitor their actual interactions with GenAI and adjust them in line with their learning goals. In this study, we use the term “metacognitive awareness” to refer to students’ perceived understanding of their own cognitive processes, including both metacognitive dimensions: knowledge of cognition and regulation of cognition in line with \textcite{Kallio2018}. 

Recent studies highlight how the unreflective use of GenAI can lead to metacognitive disengagement, a concept sometimes described as “metacognitive laziness” \parencite{Fan2025}. With GenAI offering easy and fast solutions to problems where productive struggle might be necessary for effective learning, many students turn to GenAI to get answers without having to cognitively engage, thus reducing effective learning \parencite{Zhai2024}. Conversely, productive human-AI collaboration requires students to view these tools not merely as answer generators, but as conversational partners whose contributions must be critically evaluated and meaningfully integrated into their learning processes \parencite{Benazet2026}. This necessitates developing the ability to strategically adjust human-AI collaboration in ways that are guided by their learning goals.

\subsection*{Metacognitive Interventions}

To scaffold metacognitive awareness and combat potential negative effects, recent studies highlight how an inclusion of metacognitive principles can facilitate learning in AI-assisted environments. For instance, including metacognitive planning hints was shown to increase performance in programming education \parencite{Phung2025}. Additionally, metacognitive cues increase confidence and lead to broader inquiry when using GenAI for searching \parencite{Singh2025, Singh2026}. 

However, such digital interventions lack the contextual sensitivity that human educators bring, as students learning with teacher presence have been shown to report significantly higher levels of emotional and agentic engagement than those learning without \parencite{Li2025}. Experiential and instructional alternatives have therefore emerged as complementary strategies, engaging educators in explicitly fostering metacognitive awareness through classroom interventions. Generally, experiential learning refers to the process by which knowledge and skills are acquired through direct, hands-on experience and subsequent reflection, rather than through passive instruction \parencite{Kolb1993}. On the other hand, purely instructional and lecture-based approaches describe classical structured transmission of knowledge through explicit teaching or lectures, where the educator directs learning without requiring students to engage through direct experience \parencite{Sweller2007}.

For instance, \textcite{CongLem2025} showed that experiential methods for AI-assisted even over a short workshop duration of 90 minutes were sufficient to significantly increase critical thinking in English as a foreign language in classrooms. Similarly, \textcite{Leahy2025} designed two 50-minute teacher-led interventions focused on GenAI across various higher education disciplines, finding increased student engagement with peers, instructors, and especially the subject matter. Importantly, neither of these studies compared their method to classical instructional approaches. Another example is a study by \textcite{Dickey2025} in which the authors used a primarily instructional approach with experiential elements to introduce GenAI to university students in CS courses and observed increased levels of subject-related comfort and openness. These findings highlight that structured classroom interventions, whether experiential or instructional, can meaningfully shape how students engage with GenAI in academic contexts.

However, the field still lacks a clear comparison between these approaches. To our knowledge, no study has yet compared how experiential and instructional methods influence metacognitive awareness in AI-assisted learning, nor examined the interplay between its components over time. Given the rise in students’ GenAI usage and the genuine threat of metacognitive laziness, a structured investigation into how these different approaches influence students’ metacognitive awareness is needed. Additionally, many studies call for the investigation of empirical evidence over longer time periods of the effects of metacognitive scaffolding \parencite{Fan2025, Leahy2025, Li2025}, something that this study provides.

\subsection*{The Present Study}

This short-term longitudinal study, conducted over one trimester with three measurement points, examines how to elicit metacognitive awareness on the effects of AI-assisted learning. It additionally investigates whether experiential learning approaches are more effective than purely instructional approaches for promoting such awareness and how different components of metacognitive awareness interact with each other and develop over time after a single, one time intervention. To investigate these questions, an experiential and an instructional intervention were designed matched on total instruction time. The instructional condition received a full lecture covering productive and unproductive modes of human–AI collaboration. The experiential condition received a condensed version of the same content followed by four structured exercises in which students directly enacted those interactions. The productive exercises consisted of leveraging generation effect and Socratic questioning while the unproductive interactions demonstrated cognitive offloading and failed transfer. Holding total time constant ensures that any differences between conditions reflect the mode of learning rather than exposure or time on task. 

This study builds on prior work by \parencite{Benazet2026} in which the MAI-AI instrument was developed and its constructs first explored. Whereas that initial study mapped, in an exploratory manner, the range of dimensions an intervention might shape, the present paper focuses on the metacognitive components, knowledge, and regulation of cognition, while reporting the remaining constructs for continuity and completeness. 

How best to develop students’ metacognitive awareness of GenAI effects on learning remains an open question. Lecture-based instruction has long dominated education \parencite{Stains2018}, though more active approaches such as project-based and experiential learning have gained ground in recent decades. Some researchers raise concerns that experiential learning is only effective given high initial knowledge of the taught subject \parencite{Kirschner2006, Sweller2007}, while purely instructional approaches may trigger psychological reactance \parencite{Brehm1966, Rosenberg2018} or fostering false confidence without functional awareness. Experiential learning theory suggests that concrete experience followed by reflection enables deeper understanding than instruction alone \parencite{Kolb1993, Kolb2001}. This aligns with cognitive and learning science principles suggesting that directly experiencing the consequences of productive and unproductive human-AI interactions might elicit more durable metacognitive awareness than passive instruction \parencite{Atchley2024, Kapur2008}. We thus posit the following research questions:
\begin{itemize}
\item \textbf{RQ1}: Do experiential based scaffolds lead to a higher increase in metacognitive awareness on the effects of AI-assisted learning immediately after an intervention, compared to purely instructional approaches?
\item \textbf{RQ2}: Do any gains in metacognitive awareness produced by experiential or instructional based scaffolds persist long-term after the intervention? 
\end{itemize}

\section{Materials \& Methods}

\subsection{Study Design}
\begin{figure}[!h] \includegraphics[clip,width=0.95\columnwidth]{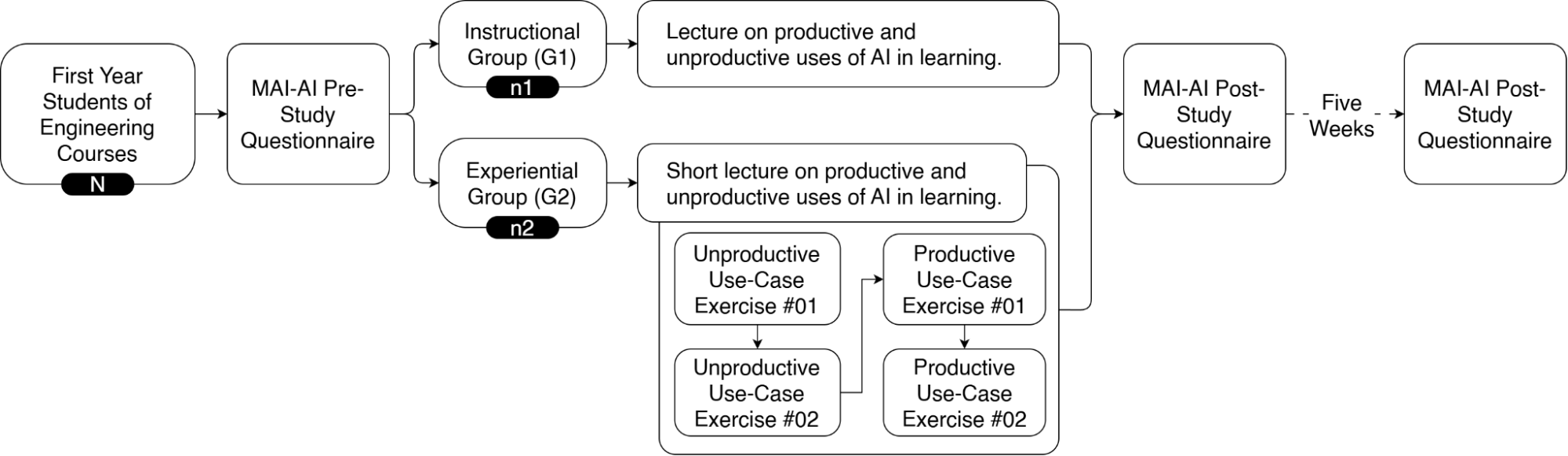}
\caption{Conceptual outline of the experimental design for initial assessment}
\label{fig:overview}
\end{figure}

As visualized in Figure \ref{fig:overview} the study was conducted using a two-group comparison design with group 1 (G1) being exposed to purely instructional guidance, while group 2 (G2) received experiential instructions consisting of productive and unproductive exercises on the same use cases. As dependent variables we measured the pre-post difference in metacognitive awareness of participants and its subcomponents, comparing their initial metacognitive awareness before and after the respective intervention. Additionally, after the initial intervention participants metacognitive awareness was assessed again after 5 weeks to track longitudinal progress. Therefore, the difference between the pre-test and the assessment after 5 weeks is recorded as an additional dependent variable. The study can thus be considered a short-term longitudinal study. Generally, the experiment used a between subject design, to ensure the absence of carry over effects. As participants were assigned to conditions based on class-schedule the study concerns a quasi-experimental design. 

\subsection{Participants}

To determine the necessary sample size, an a priori power analysis was run in GPower \parencite{Faul2007}. Assuming a medium effect ($d$ = .5), $\alpha$ = .05, and power = .80 for independent t-tests, a minimum of 128 participants was required.

Overall, N=138 first-year engineering students from a Universitat Pompeu Fabra, Barcelona were recruited for the experiment, which took part through an authentic  course activity. Data was considered only for those who voluntarily consented to share it for research purposes (see Ethical Considerations below). Participants were assigned either to the experiential group ($n_2$=69) or the instructional group  ($n_1$=69) based on randomly assigned class schedules with the sole inclusion criteria of being at least 18 years old to comply with ethical guidelines. As the study was seamlessly integrated within the regular class schedule no explicit incentive was provided for participation and students declining participation were taught on a separate date. 

\subsection{Materials}

\subsubsection*{Learning Task \& Interventions}

The learning task at hand centred around the question of how to effectively collaborate with GenAI during learning. In the instructional group, participants received a traditional lecture in which productive and unproductive GenAI partnership use cases were explained didactically. They listened to examples and explanations without any hands-on engagement. In the experiential group, rather than passively receiving lecture content, participants engaged in four structured exercises that experientially demonstrated the concepts. As productive and unproductive use cases were briefly introduced, participants immediately practiced them through two productive GenAI collaboration scenarios and two unproductive GenAI collaboration scenarios. The experiential productive and unproductive exercises can be found in the additional material. 

Generally, all experiential exercises for the experiential group followed a similar structure: (1) participants perform a task with the help of a GenAI system (approx. 5-10 min), (2) afterwards, participants performs a distraction task, on an unrelated topic (approx. 5 min), (3) finally participant perform a similar task related to the initial one, this time without the help of any GenAI systems (approx. 10 min). The exact productive and unproductive exercises are described in Table \ref{tab:learning_task}. 

Our experiential condition, is structured and scaffolded practice; students receive condensed direct instruction before each exercise, and additionally the exercise specifies the task, the form of GenAI interaction, and a built-in reflection step (see Table \ref{tab:learning_task}), providing the guidance novice learners require while still grounding learning in direct experience \parencite{Kirschner2006, Sweller2007}.

\begin{table*}[h!]
\centering
\caption{Description of productive and unproductive exercises}
\label{tab:learning_task}
\begin{tabular}{p{8cm}|p{8cm}}
\hline
\textbf{Productive} & \textbf{Unproductive}
\\
\hline
Productive Exercise 1 \newline
\textbf{``The Generation Effect Activator''} \newline
(1) Reading a technical article and writing an initial 50--100-word summary without GenAI,\newline
(2) submitting the summary to a GenAI model with the prompt ``Here's my summary [paste]. What key points did I miss or misunderstand?'',\newline
(3) revising the summary based on GenAI feedback,\newline
(4) after distraction tasks, applying concepts to five novel scenarios. \newline
\newline
This exercise applies the generation effect \parencite{Bertsch2007, Slamecka1978} and productive failure principles \parencite{Kapur2008, Kapur2012}. Also it delays GenAI use until after initial memory encoding, which may support memory formation \parencite{Kosmyna2025}. & 
Unproductive Exercise 1 \newline
\textbf{``The Problem Solution Bypass''} \newline
(1) Receiving a computational thinking problem,\newline
(2) immediately requesting a complete solution from GenAI,\newline
(3) copying the solution,\newline
(4) following distraction, attempting a similar transfer problem independently.\newline
\newline
This design removes the struggle necessary for learning \parencite{OHara1998}, contradicting productive failure principles \parencite{Kapur2008}, and encourages cognitive offloading \parencite{Grinschgl2021, Sparrow2011} by preventing active problem-solving. \\
\hline
Productive Exercise 2 \newline
\textbf{``The Socratic AI Challenger''} \newline
(1) Writing three arguments supporting a position on a controversial statement,\newline
(2) engaging with a GenAI model using the prompt ``Act as a Socratic tutor on [topic]. Challenge my arguments with questions only: [arguments]'',\newline
(3) refining arguments based on the exchange,\newline
(4) after distraction tasks, generating three counterarguments with defenses independently.\newline
\newline
This exercise uses GenAI as a Socratic tutor which was shown to be a productive collaborative framework \parencite{Elder1998, Ho2023, Orynbassarova2024}. 
& 
Unproductive Exercise 2 \newline
\textbf{``The Passive Summary Trap''} \newline
(1) Receiving a complex academic article,\newline
(2) requesting a GenAI-generated bullet-point summary,\newline
(3) reviewing the summary passively,\newline
(4) following distraction, explaining core concepts in simplified terms appropriate for a non-expert audience.\newline
\newline
This design theoretically produces shallow encoding \parencite{Kosmyna2025} and impairs semantic retention \parencite{Abbas2024}, preventing development of transferable conceptual understanding. \\
\hline
\end{tabular}
\end{table*}

\subsubsection*{MAI-AI}

To assess the research questions raised by this study the Metacognitive Awareness Inventory for Artificial Intelligence (MAI-AI) was used \parencite{Benazet2026}. This questionnaire is an AI-centered adaptation of the shortened version of the Metacognitive Awareness Inventory (MAI) \parencite{GonzalezCabanes2022, Harrison2017, Schraw1994} and the Motivated Strategies for Learning Questionnaire (MSLQ) \parencite{Duncan2005}.

Overall the MAI-AI questionnaire consists of 29 questions (28 in post), divided into 6 constructs: Attitudes towards AI (Attitudes - AIA), current and future use of AI systems (AI Engagement - AIE), knowledge on how AI systems function (AI Literacy - AIL), metacognitive knowledge of the effects of human-AI collaboration on learning processes (Metacognitive Knowledge of Cognition - MKC), strategic regulation on human-AI collaboration in learning processes (Metacognitive Regulation of Cognition - MRC) and the feeling of ownership and control over their work (Autonomy \& Control - ANC). The pre-study questionnaire includes one extra question regarding current AI system usage on academic work (Q3). The full table with the complete MAI-AI Questionnaire can be found in the additional material.

All questions are answered on a five point Likert scale with the average regarding subscales and the whole scale being considered the score. Therefore, the final scores could range from 1 to 5. 

\subsection{Procedure}

The study was conducted in two two-hour sessions led by the researchers. They both followed a standardized protocol specifying identical content scope, scripted explanations, and fixed time allocations for each segment and task, with similar content but different instructional methods. While instructors thus knew which group was considered experimental, participants were blinded to this information. All participants first completed the pre-test MAI-AI questionnaire after which both groups received instruction on productive and unproductive AI-assisted learning patterns in educational contexts through different pedagogical approaches.

While the instructional group received a traditional lecture in which productive and unproductive GenAI partnership use cases were explained didactically, the experiential group, rather than passively receiving lecture content, engaged in four structured exercises that experientially demonstrated the concepts shortly introduced beforehand. In the experiential group, participants had unrestricted access to Google Gemini 2.5 during the four exercises. Following the intervention phase, both groups completed the post-test MAI-AI questionnaire. All participants spent the full two hours within their respective session. To track longitudinal progress of metacognitive awareness in relation to learning with GenAI both groups redid the MAI-AI questionnaire exactly five weeks after the initial session. 

\subsection{Data Analysis Plan}
Initial cleaning removed all participants who failed to complete both the pre-test and post-test questionnaires, reducing the sample to $N$ = 128. Subsequently, to ensure data quality, all responses were screened for careless responding patterns. Three indicators were flagged: straight-lining (identical answer for $>$75\% of questions), extreme responding ($>$75\% using 1s or 5s) and low variance (SD $<$ 0.5). Participants flagged for two or more indicators were automatically excluded. Participants flagged for straight-lining alone were also excluded. This process yielded a final sample of $N$ = 126 ($n_1$ = 61, $n_2$= 65), retaining 91.3\% of the initial responses. Additionally, participants failing to complete the MAI-AI after 5 weeks were excluded from the analysis of RQ2, which resulted in a final sample of $N$ = 107 ($n_1$ = 57, $n_2$= 50). This accounts for the slight differences in mean and standard deviation between the post immediate and post five weeks analysis.

For each construct, normality of the change scores was assessed via the Shapiro–Wilk test; paired (within-group) or independent (between-group) t-tests were used where normality held, and Wilcoxon signed-rank or Mann–Whitney U tests otherwise. 

To control the false discovery rate across the six constructs within each comparison family (within-group and between-group, at each timepoint), we applied the Benjamini–Hochberg correction as our primary control. Bonferroni-corrected results are additionally reported as a conservative benchmark. 

\subsection{Ethical Considerations}
The study received ethical approval from the Universitat Pompeu Fabra ethics review body (CIREP; approval no. 452). All participants provided informed consent prior to taking part and were informed of their right to withdraw at any time without providing a reason. Data was stored in pseudo-anonymized form until the last session after which it was fully anonymized and stored on secure university servers in accordance with GDPR guidelines.

\section{Results}

MAI-AI scores were collected at three different points: before the intervention (Pre - T0) immediately after the intervention (Post Immediate - T1) and five weeks after the intervention (Post Five Weeks - T5). Construct-level analyses have been conducted between the two latter measurement points and the pre-intervention MAI-AI scores, and within and between the two groups (Group 1 - Instructional, Group 2 - Experiential). The full analysis and data set can be found in the additional material.

Because the immediate post-test and five-week follow-up analyses were conducted on partially different samples (owing to attrition and the independent careless-responder screening applied at each stage; $N$ = 126 and $N$ = 107, respectively), baseline equivalence between the control and intervention groups was verified separately for each analytic sample. For every construct, pre-test scores were compared using an independent-samples t-test or, where the normality assumption was violated, a Mann–Whitney U test. No statistically significant baseline differences emerged for any construct in either sample (all $p >$ .05), indicating that the groups were statistically equivalent at baseline in both analyses.

\subsection{Pre-Post Immediate Change}

\subsubsection*{Within-groups}
In the instructional group (G1), three constructs showed significant improvement on pre-post change between T0 and T1. Attitudes (AIA) shifted from $M$ = 3.43 ($SD$ = 0.76) to 3.70 ($SD$ = 0.67) (Wilcoxon $p <$ .001, $r$ = .91); AI Literacy (AIL) from 3.64 ($SD$ = 0.69) to 3.86 ($SD$ = 0.61) ($t$(61) = 3.73, $p <$ .001); and AI Engagement (AIE) from 3.66 ($SD$ = 0.44) to 3.83 ($SD$ = 0.42) (Wilcoxon $p <$ .001, $r$ = .53). Metacognitive Regulation of Cognition (MRC) ($\Delta M$ = 0.11, $p$ = .051, $d$ = 0.26), Metacognitive Knowledge of Cognition (MKC) ($\Delta M$ = 0.05, $p$ = .42) and Autonomy \& Control (ANC) ($\Delta M$ = 0.11, $p$ = .31) did not show any significant difference from pretest scores.

In the experiential condition (G2), three constructs likewise showed significant changes. AIE  increased substantially, from $M$ = 3.60 ($SD$ = 0.45) to 4.01 ($SD$ = 0.50) ($t$(65) = 8.26, $p <$ .001, $d$ = 1.00); MKC showed a comparable reliable gain, from 3.57 ($SD$ = 0.64) to 3.95 ($SD$ = 0.63) (Wilcoxon $p <$ .001, $r$ = .78) and AIL also improved, from 3.49 ($SD$ = 0.63) to 3.62 ($SD$ = 0.70) (Wilcoxon $p$ = .004, $r$ = .65). MRC ($\Delta M$ = 0.08, $p$ = .058), AIA ($\Delta M$ = 0.13,$p$ = .15) and ANC ($\Delta M$ = 0.02, $p$ = .97) were not reliably altered.

\subsubsection*{Between-groups}

Significant differences on pre-post change scores between T0 and T1 were found between groups on two constructs. The intervention favoured AIE ($U$ = 1242, $p <$ .001, $r$ = .37), with the experiential group (G2) exhibiting a change more than twice that of the instructional group ($\Delta M$ = +0.41 vs. +0.18). MKC was also significantly increased by the intervention ($U$ = 1315.5, $p$ = .001, $r$ = .34), with a substantive gain (+0.38) against a near-flat instructional group trajectory (+0.05). The remaining four constructs yielded no significant between-group differences: AIA ($p$ = .25), AIL ($p$ = .55), MRC ($p$ = .87) and ANC ($p$ = .42).

\begin{figure*}[h!]
    \centering
    \begin{minipage}{0.48\textwidth}
        \centering
        \includegraphics[width=\linewidth]{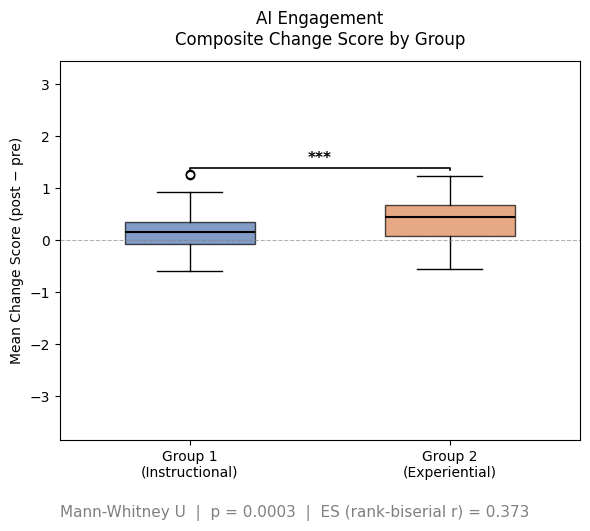}
    \end{minipage}
    \hfill
    \begin{minipage}{0.48\textwidth}
        \centering
        \includegraphics[width=\linewidth]{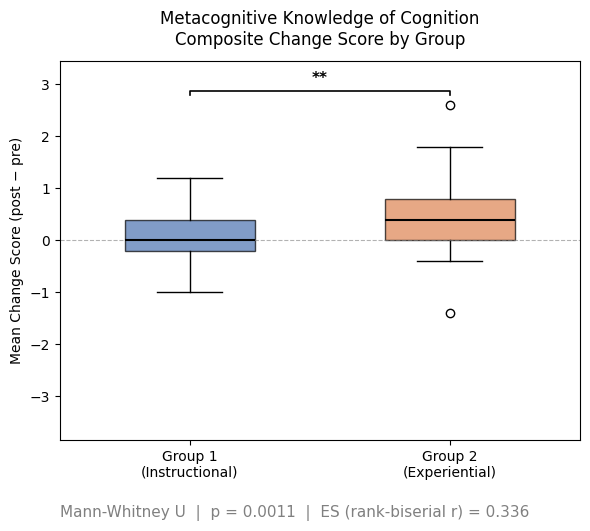}
    \end{minipage}
    \caption{Significant change scores differences between groups in AIE and MKC at T1 - Post Immediate.}
    \label{fig:between_immidiate}
\end{figure*}

\subsection{Pre - Post Five Weeks Change}

\subsubsection*{Within-groups}

In the instructional group (G1), of the six constructs, only one survived multiple-comparison corrections between T0 and T5: AI Engagement (AIE) showed a substantial sustained gain ($M_{pre}$ = 3.66, $M_{postW5}$ = 4.01; $\Delta M$ = +0.35; Wilcoxon $p <$ .001, $r$ = .72; Bonferroni $p <$ .001, BH $p <$ .001). All other constructs did not display significant changes. The instructional group was therefore dominated by a single construct improving: AI Engagement (AIE), while all other constructs remained essentially flat.

In the experiential group (G2) three constructs survived at least BH correction. AIE showed a large sustained gain comparable to that of the instructional group ($\Delta M$ = +0.30; Wilcoxon $p <$ .001, $r$ = .73; Bonferroni $p <$ .001, BH $p <$ .001). MRC improved reliably ($\Delta M$ = +0.16; $p$ = .002, $d$ = 0.37; BH $p$ = .048). MKC showed a medium-sized gain ($\Delta M$ = +0.25; $p$ = .006, $d$ = 0.35; BH $p$ = .048). AIL ($\Delta M$ = +0.12, $p$ = .117), AIA ($\Delta M$ = +0.11, $p$ = .182) and ANC ($\Delta M$ = +0.09, $p$ = .448) did not reach significance.

\subsubsection*{Between-groups}

Direct between-group comparisons on change scores yielded no contrasts that survived multiple-comparison correction. The only construct comparison that approached significance was MKC, where the experiential group's gain (+0.25) was magnitudes higher than the instructional's loss (-0.003), yielding $t$(106) = -1.94, $p$ = .055, $d$ = -0.37 (medium effect, favouring the experiential group). This is not significant at conventional thresholds and does not survive multiple-comparison correction, but the effect size is substantively meaningful.

\begin{figure*}[h!]
    \centering
    \begin{minipage}{0.48\textwidth}
        \centering
        \includegraphics[width=\linewidth]{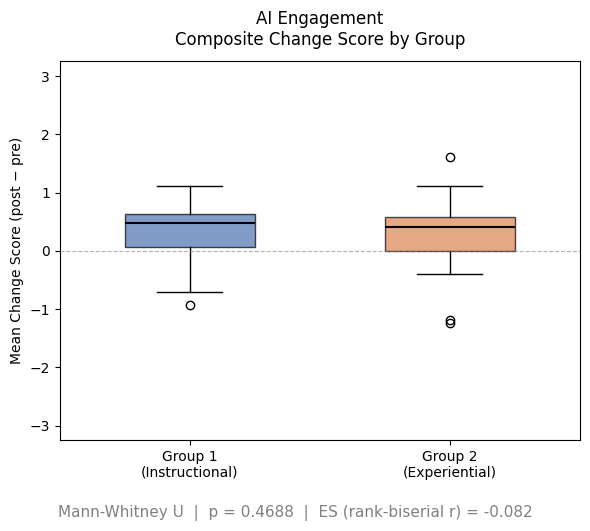}
    \end{minipage}
    \hfill
    \begin{minipage}{0.48\textwidth}
        \centering
        \includegraphics[width=\linewidth]{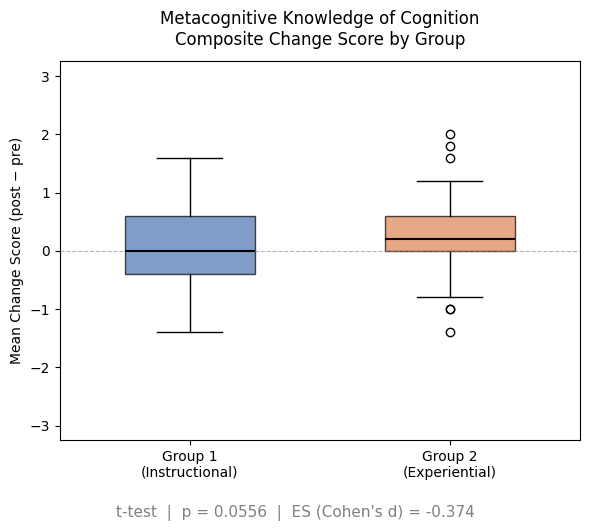}
    \end{minipage}
    \caption{Change scores between groups in AIE and MKC at T5 - Post Five Weeks.}
    \label{fig:between_5_weeks}
\end{figure*}

\subsection{Pre - Post Immediate - Post Five Weeks Evolution}

To complement the previous analyses, evolution plots were produced for each of the six construct composites, displaying group means at Pre (T0), Post Immediate (T1), and Post Five Weeks (T5). These plots aid in visualising the shape of the within-condition trajectories. Additionally, a descriptive overview is provided in Table \ref{tab:means_sd}

\begin{figure}[h!]
    \centering
    \begin{minipage}{0.48\textwidth}
        \centering
        \includegraphics[width=\linewidth]{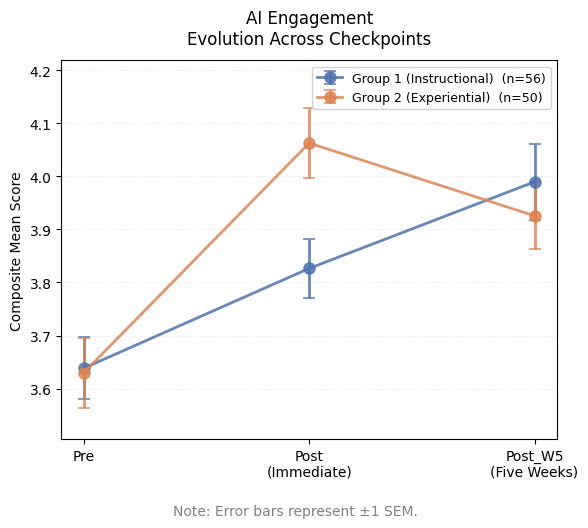}
    \end{minipage}
    \hfill
    \begin{minipage}{0.48\textwidth}
        \centering
        \includegraphics[width=\linewidth]{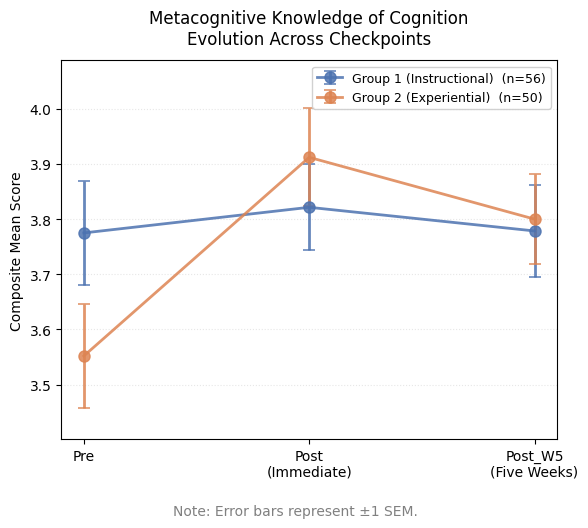}
    \end{minipage}
    
    \vspace{0.5cm}
    
    \begin{minipage}{0.48\textwidth}
        \centering
        \includegraphics[width=\linewidth]{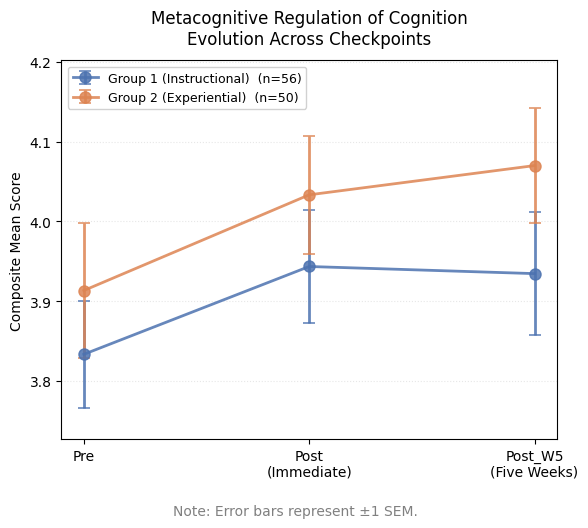}
    \end{minipage}
    
    \caption{Evolution plots for test results in AIE, MKC and MRC at Pre, Post Immediate and Post Five Weeks}
    \label{fig:evolutions}
\end{figure}

\subsubsection*{AI Engagement}
    The instructional group (G1) shows monotonic growth across all three checkpoints (3.64 $\rightarrow$ 3.82 $\rightarrow$ 3.99), whereas the experiential group spikes sharply at Post Immediate (3.63 $\rightarrow$ 4.06) and then regresses slightly by Post Five Weeks (3.93). Both groups arrive at a comparable level of self-reported AI engagement five weeks after the intervention, but the paths there are qualitatively different: the instructional group acquires engagement gradually, while the intervention produces a noticeable early peak that is retained long term.

\subsubsection*{Metacognitive Knowledge of Cognition (MKC)}
    The instructional group is essentially flat across all three checkpoints (3.78 $\rightarrow$ 3.82 $\rightarrow$ 3.78), whereas the intervention shows a sharp gain at Post-Immediate (3.55 $\rightarrow$ 3.91) followed by a partial regression (3.80). The experiential group (G2) starts at a numerically lower baseline and has a significant increase at T1 (Post Immediate), maintaining such gain at T5 (Post Five Weeks). 

\subsubsection*{Metacognitive Regulation of Cognition (MRC)}

    This trajectory pair, although not displaying any significant difference in change scores, is visually consistent with a delayed-divergence pattern, and warrants particular attention given the importance of MRC to the intervention’s theoretical rationale. The instructional group (G1) shows a small immediate gain that is followed by a plateau. In contrast, the experiential group (G2) displays an immediate gain which is followed by continued growth. 

\begin{table}[h!]
\centering
\begin{tabular}{llccc}
\toprule
\textbf{Construct} & \textbf{Group} & \textbf{$T_0$ - $M$ ($SD$)} & \textbf{$T_1$ - $M$ ($SD$)} & \textbf{$T_5$ - $M$ ($SD$)} \\
\midrule
\multirow{2}{*}{AIA} & G1 (Instructional) & 3.43 (0.78) & 3.70 (0.68) & 3.61 (0.78) \\
                     & G2 (Experiential)  & 3.38 (0.82) & 3.61 (0.77) & 3.49 (0.78) \\
\midrule
\multirow{2}{*}{AIE} & G1 & 3.64 (0.44) & 3.82 (0.42) & 3.99 (0.53) \\
                     & G2 & 3.63 (0.47) & 4.06 (0.46) & 3.93 (0.44) \\
\midrule
\multirow{2}{*}{AIL} & G1 & 3.60 (0.69) & 3.84 (0.62) & 3.68 (0.62) \\
                     & G2 & 3.43 (0.62) & 3.60 (0.67) & 3.55 (0.63) \\
\midrule
\multirow{2}{*}{MKC} & G1 & 3.78 (0.70) & 3.82 (0.58) & 3.78 (0.63) \\
                     & G2 & 3.55 (0.66) & 3.91 (0.63) & 3.80 (0.58) \\
\midrule
\multirow{2}{*}{MRC} & G1 & 3.83 (0.50) & 3.94 (0.53) & 3.94 (0.58) \\
                     & G2 & 3.91 (0.60) & 4.03 (0.52) & 4.07 (0.51) \\
\midrule
\multirow{2}{*}{ANC} & G1 & 3.74 (0.66) & 3.82 (0.71) & 3.89 (0.62) \\
                     & G2 & 3.66 (0.71) & 3.61 (0.67) & 3.75 (0.69) \\
\bottomrule
\end{tabular}
\caption{Means and Standard Deviations at the three measure points for each construct and group.}
\label{tab:means_sd}
\end{table}

\section{Discussion}

With AI-assisted learning gaining increasing importance in modern day education the question of which teaching approaches best facilitate metacognitive awareness, a skill central to learning, becomes pressing. Current research is lacking a clear and structured comparison of how experiential and instructional methods in the classroom affect long-term metacognitive awareness in AI-assisted learning, as well as the interactions among its components. Our results indicate that experiential methods generally take the upper hand. Immediately after learning, students receiving experiential instructions showed higher levels of knowledge of cognition and general engagement with GenAI, which persisted throughout the semester. At the same time instructional methods did not lead to immediate change in metacognitive awareness, but reached similar levels of knowledge and engagement later on. Additionally, a delayed increase in regulation of cognition in the experiential group was observed, raising both important questions about the desirability of increased engagement without MRC and showing promising long-term effects. 

RQ1 concerned the effects of the two approaches on metacognitive awareness. Our results suggest that experiential methods outperform instructional approaches immediately after the intervention (T0-T1 comparison), with both engagement and MKC increasing significantly more than in the instructional group. These findings are in line with previous research in non-AI contexts showcasing that experiential instructions lead to higher metacognitive awareness in general when compared to instructional approaches \parencite{Akkas2021, Ho2025}. However, our results highlight that these gains are more nuanced and can diverge strongly across dimensions of metacognitive awareness. For one a rise in knowledge of cognition was observed which reflects a greater understanding of how GenAI shapes one's own learning. Plausible reasons for this relate back to experiential exposure making the consequences of productive and unproductive GenAI use directly observable in a way instruction cannot \parencite{Fan2025, Kapur2008, Kolb1993}.

Notably, the increase in knowledge of cognition was not accompanied by a significant rise in regulation of cognition, the insight about translating that knowledge into reflective, controlled practice. This dissociation is consistent with theoretical accounts treating knowledge and regulation as separable layers, in which monitoring does not necessarily produce effective control \parencite{Efklides2014}. Additionally, empirical findings highlight that knowledge of one's strategies does not guarantee their use with possible reasons for missed transfer being cost-benefits judgements \parencite{Yamaguchi2023}. Combined with the finding that experiential learning increased engagement and thus GenAI use, this raises concerns about whether greater use without reflective action may in itself lead to cognitive offloading \parencite{Gerlich2025, Zhai2024}. This might diminish some positive aspects of experiential learning and paints a more nuanced picture of its advantage over instructional approaches. 

Regarding long-term effects explored in RQ2, our findings indicate that the initial gap between experiential approaches and instructional methods in terms of knowledge of cognition and AI engagement closes over time with both groups showing comparable increased levels at the end of the semester. This pattern is consistent with the well-documented phenomenon of intervention “fade-out”, which most often takes the form of a control-group catch-up rather than an absolute decline in the treatment group \parencite{Bailey2020}.  Even though literature on long term effects is thin \parencite{Fan2025, Leahy2025, Li2025}, it is possible that prolonged and individual experimentation in the instructional group lead to a delayed increase in engagement and knowledge, resulting in similar levels as the experiential group. This can be interpreted as the experiential intervention accelerating early engagement that would otherwise develop slowly through ordinary exposure to GenAI tools and that a single intervention without consistent experiential sessions is insufficient to continually increase knowledge of cognition \parencite{Hargrove2013}. 

While knowledge and engagement converged across groups, exploratory analyses revealed a different trajectory for regulation of cognition. The experiential group showed a continuous within-group increase across the three measurement points, that was not observed in the instructional group. While the instructional group plateaued in the measurement after 5 weeks the experiential group elicited a significant increase in regulation, which should however not be confused with a statistical between-group advantage \parencite{Gelman2006}. While the relationship between knowledge and regulation is generally seen as complex and mutually reinforcing \parencite{Schraw1994}, the experiential learning cycle offers an explanation for this phenomenon \parencite{Kolb1993}. It posits that learning is a cyclical process in which experience is translated into abstract conceptualization and regulation through reflection over time rather than instantaneously. In this light, the knowledge gained immediately after the intervention may have functioned as the basis that students gradually converted into reflective regulation. This suggests that the immediate knowledge gain and the delayed regulation gain are part of a single developmental process rather than independent outcomes under experiential learning. However, since the between-group difference did not reach statistical significance, possibly due to a lack of power, these findings warrant future exploration about whether robust statements can be made over the long-term advantage of experiential approaches in terms of regulation of cognition. 

\section{Implications}
This study offers one of the first systematic comparisons of experiential and instructional teaching approaches in AI-assisted learning. Experiential instruction produced larger immediate gains in knowledge of cognition and engagement with GenAI, alongside a within-group increase in regulation of cognition that emerged over time. Although instructional approaches eventually reached comparable levels of knowledge and engagement, this convergence took weeks to unfold. Practitioners are thus advised to primarily rely on experiential and hands-on approaches for cultivating metacognitive awareness in AI-assisted learning.

Theoretically, our results suggest that components of metacognitive awareness are not only separable but can be temporally decoupled under experiential AI-assisted learning. This has strong methodological implications. A null result for regulation immediately after an intervention may reflect delayed development through experiential reflection rather than a genuine absence of effects. Design relying solely on post-test measurements thus risks misclassifying regulatory effects that only emerge over time. 

Our results also surface a tension. Experiential instruction raised both engagement and knowledge of cognition without a corresponding immediate rise in regulation, leaving a gap between students' understanding of effective strategies and the knowledge of how to apply them. Because greater engagement without regulation may itself lead to cognitive offloading, simply encouraging more GenAI use is unlikely to close this gap. Due to this gap between understanding of strategies and their translation into practice, practitioners should not get frustrated when experiential exercises do not lead to immediate regulated practice but require some time to unfold. They should rather offer spaces for independent use in which students can refine their regulatory behaviour through repeated cycles of use and structured reflection.

\section{Limitations \& Future Work}
Despite the contributions of this study, certain methodological and contextual limitations must be considered when evaluating the results and their broader applicability. First, metacognition was assessed via repeated self-reported administrations of the same questionnaire. While questionnaires are the most commonly used instrument to assess metacognition they inherently suffer from limitations related to subjective reporting, including social desirability bias and respondents' varying ability to accurately evaluate their own cognitive processes \parencite{Boekaerts2005, Cromley2007, Harrison2017}. Additionally, repeated exposure to the same instrument may have increased participants' familiarity with the questionnaire items, potentially influencing responses independently of actual changes in metacognitive awareness. Finally it has to be noted that the MAI-AI instrument in question is still awaiting validation. Future research could potentially include more varied assessment methods such as think-aloud protocols or performance-based measures, alongside self-report questionnaires \parencite{Dinsmore2008, Winne2012}.

Second, there was a lack of environment variables control since students could use GenAI as personally deemed useful between the initial session and the final measurement point. While it is difficult to control for this kind of confound, future research should nonetheless try to actively monitor whether differences in independent use between groups could have disproportionately influenced long-term results. 

Third, there was no demographic data collected to illustrate approximately equal distribution between groups and allow for exploration of age or gender effects. Future research should systematically capture these attributes to strengthen the validity of the results. 

Fourth, the final analysed sample includes $N$ = 126 participants, falling somewhat short of the 128 participants required by the a priori power analysis to detect medium effects. The study was thus slightly underpowered for medium between-group effects, so the non-significant MRC difference at five weeks may reflect limited power rather than a true absence of different development and should be confirmed in future research. 

Finally, the sample was drawn exclusively from first-year engineering students at Universitat Pompeu Fabra, Barcelona. As a result, the findings may not generalize to students from other disciplines, educational levels, or cultural contexts. Future work should replicate the study across more diverse student populations, especially in secondary education where GenAI is a new and disrupting phenomenon, to establish the robustness and generalizability of the observed effects.

\section{Conclusion}
We present one of the first systematic comparisons of experiential and instructional teaching approaches on metacognition under AI-assisted learning. In a short-term longitudinal study we monitor first-year undergraduate students' metacognitive development in regard to reflective practices, following an either experiential or instructional, lecture-based intervention. Our results indicate that experiential methods outperform instructional approaches in terms of engagement and metacognitive knowledge of cognition immediately after the intervention. Additionally, the experiential group showed a delayed but continuous within-group increase in regulation of cognition not seen in the instructional group. This suggests that beneficial scopes of experiential methods extend beyond traditional lecturing towards AI-assisted learning. The central challenge may thus lie in designing interventions that not only spark immediate awareness but sustain its slower-developing regulatory counterpart over time.

\section*{Acknowledgments}

\subsection*{Funding}
This work was supported by the Spanish Research Agency (AEI) under PID2023-146692OB-C33 and the Instituci\'{o} Catalana de Recerca i Estudis Avan\c{c}ats. 

\subsection*{Additional Material}

Additional material can be found on Zenodo under \\\url{https://zenodo.org/records/21234837?token=eyJhbGciOiJIUzUxMiJ9.eyJpZCI6IjY2M2U4NjEyLTkzZjAtNDgzYS05NzdkLTllN2RmZTIwNzExMyIsImRhdGEiOnt9LCJyYW5kb20iOiI0Y2Y3ZGM3MTJhM2ZkZDhkMDU0ODRmMzc3ZjJiNWE1ZSJ9.SriaRqEK3RMPoWVVVU6Wo2gDtKvWAgDhP9FbNXjBOPoXRg2vBMOfv4Qmefw9AXg8TROi97Ow4btgIHDsi2XPQA}.

\subsection*{Generative AI Use}
Analysis scripts for statistical tests and the Jupyter notebook structures were generated with the assistance of Claude Code (Sonnet 4.6 and Opus 4.8) inside a Visual Studio Code environment (last accessed April 2026). All AI-generated code was reviewed, tested, and revised by the authors, who confirmed accuracy and reproducibility of results. The authors take full responsibility for it.


\printbibliography



\end{document}